# Hybrid graphene/silicon integrated optical isolators with photonic spin–orbit interaction

Jingwen Ma,[1,2] Xiang Xi,[1] Zejie Yu,[1] and Xiankai Sun[1,2,a]

[1]*Department of Electronic Engineering, The Chinese University of Hong Kong, Shatin, New Territories, Hong Kong*
[2]*Shun Hing Institute of Advanced Engineering, The Chinese University of Hong Kong, Shatin, New Territories, Hong Kong*

**Abstract**
Optical isolators are an important building block in photonic computation and communication. In traditional optics, isolators are realized with magneto-optical garnets. However, it remains challenging to incorporate such materials on an integrated platform because of the difficulty in material growth and bulky device footprint. Here, we propose an ultracompact integrated isolator by exploiting graphene's magneto-optical property on a silicon-on-insulator platform. The photonic nonreciprocity is achieved because the cyclotrons in graphene experiencing different optical spin exhibit different response to counterpropagating light. Taking advantage of cavity resonance effects, we have numerically optimized a device design, which shows excellent isolation performance with the extinction ratio over 45 dB and the insertion loss around 12 dB at a wavelength near 1.55 µm. Featuring graphene's CMOS compatibility and substantially reduced device footprint, our proposal sheds light to monolithic integration of nonreciprocal photonic devices.

---

[a]Author to whom correspondence should be addressed. Electronic mail: xksun@cuhk.edu.hk

Optical isolators are a type of nonreciprocal devices that allow for unidirectional light transmission by breaking the time-reversal symmetry.[1,2] They are an important building block in photonic computation and communication systems because of efficient suppression of the undesired back reflection.[3] In traditional optics, such devices can easily be implemented by utilizing materials with Faraday rotation effect. However, it is difficult to incorporate them into an integrated platform because of large lattice mismatch and thermal incompatibility between the magneto-optical garnet and the substrate.[4] To date, much effort has been focused on heterogeneous wafer bonding techniques[5-8] or depositing CMOS-compatible magneto-optical materials.[9-11] Nevertheless, with device size at least hundreds of micrometers, most devices are too cumbersome for upscaling and mass production. On-chip integrated isolators with a suitable material, compact size, and decent performance are still desperately desired.

Graphene, a monolayer of carbon atoms, has recently attracted intense interest in integrated photonics for its CMOS compatibility and unique electronic band structures. Various devices such as photodetectors,[12,13] optical modulators,[14,15] and LEDs[16] have been reported exploiting its characteristics of high electron mobility and tunable broadband light–matter interaction.[17-19] On the other hand, graphene also exhibits enormous magneto-optical effect.[20] Recent experimental results have shown that magnetically biased graphene can break the time-reversal symmetry for propagating light, causing a Faraday rotation in its polarization state when the optical spin is perpendicular to the plane of carbon atoms.[21,22] This effect can be utilized to construct optical isolators. To date, most graphene-based optical isolators take an out-of-plane scheme, where light propagates along the normal of the graphene sheet.[23-25] This is because light in an infinite medium is a transverse wave with its spin in parallel with the propagation direction. Such an out-of-plane scheme not only suffers from the short interaction length, but also is difficult for on-chip implementation where light usually propagates in a plane in parallel with the graphene sheet.[12-15]

In this Letter, we propose a hybrid graphene/silicon magneto-optical isolator for on-chip integration that works in the fiber-optic communication band. By exploiting the spin–orbit interaction[26,27] in nanophotonic structures together with graphene's magneto-optical effect, our proposed isolator can achieve excellent isolation performance with the extinction ratio as high as 45 dB and the insertion loss around 12 dB at the wavelength of 1.552 μm. The excellent device performance, together with the ultracompact size and the advantage of CMOS compatibility, has shown its great promise in on-chip integration of photonic nonreciprocal devices.

As shown in Fig. 1(a), the optical isolator consists of a photonic bus waveguide and a microring resonator, both fabricated on a silicon-on-insulator substrate. A patterned graphene nanoribbon covers the inner top surface of the silicon waveguide of the microring. With an external magnetic field perpendicularly applied to the device plane, we expect distinct light transmission spectra for the two opposite propagation directions, thus achieving the function of optical isolation. The operation mechanism of isolation is shown in Fig. 1(b): owning to the photonic spin–orbit interaction,[26-30] the magnetically induced cyclotrons in the graphene nanoribbon experience distinct photonic spin for light of opposite propagation directions. As a result, graphene's magneto-optical property induces a difference in the effective refractive index of the forward and backward propagating light, causing nonreciprocal transmission spectrum required for an optical isolator.

The nonreciprocal device performance is closely related to graphene's optical properties. In a vertically applied magnetic field, graphene's energy band splits into discrete Landau levels and its optical conductivity **σ** can be described by a tensor with both longitudinal $\sigma_L$ and Hall $\sigma_H$ components:

$$\boldsymbol{\sigma} = \begin{pmatrix} \sigma_L & \sigma_H \\ -\sigma_H & \sigma_L \end{pmatrix} \tag{1}$$

For photons with a spin state $(1, \pm i)^T$, their effective optical conductivity can immediately be determined to be $\sigma_L \pm i\sigma_H$. One can see that the nonzero Hall conductivity term $\sigma_H$ causes different optical response for the right-and left-handed circularly polarized light.[24,31,32] By the Kubo method, the longitudinal and Hall components are expressed as follows:

$$\sigma_L = \frac{e^2 v_F^2 |eB_0|(\omega - j2\Gamma)\hbar}{j\pi}$$
$$\times \sum_{n=0}^{\infty} \left[ \frac{(f(E_{i+1}) - f(E_i)) + (f(-E_i) - f(-E_{i+1}))}{(E_{i+1} - E_i)^2 - (\omega - j2\Gamma)^2 \hbar^2} \times \frac{1 - \Delta^2/E_i E_{i+1}}{E_{i+1} - E_i} \right.$$
$$\left. + \frac{(f(E_{i+1}) - f(-E_i)) + (f(E_i) - f(-E_{i+1}))}{(E_{i+1} + E_i)^2 - (\omega - j2\Gamma)^2 \hbar^2} \times \frac{1 + \Delta^2/E_i E_{i+1}}{E_{i+1} + E_i} \right] \quad (2)$$

$$\sigma_H = \frac{e^2 v_F^2 eB_0}{\pi} \times \sum_{n=0}^{\infty} \left[ (f(E_{i+1}) - f(-E_i)) - (f(E_i) - f(-E_{i+1})) \right]$$
$$\times \left[ \frac{1 - \Delta^2/E_i E_{i+1}}{(E_{i+1} - E_i)^2 - (\omega - j2\Gamma)^2 \hbar^2} + \frac{1 + \Delta^2/E_i E_{i+1}}{(E_{i+1} + E_i)^2 - (\omega - j2\Gamma)^2 \hbar^2} \right] \quad (3)$$

where $f(E) = \{\exp[(E - \mu)/k_B T] + 1\}^{-1}$ is the Fermi–Dirac distribution, $E_i = (\Delta^2 + 2i|eB_0|\hbar v_F^2)^{1/2}$ denotes the $i$-th Landau energy, $T$, $k_B$, $e$, $c$, and $\hbar$ are respectively the temperature, the Boltzmann constant, the electron charge, the speed of light in vacuum, and the reduced Planck constant. Under an extremely strong magnetic field $B_0$, an energy gap $\Delta$ may form in graphene's band structure. However, with typical values of $B_0$ discussed in this work, $\Delta$ is negligibly small and can be regarded as zero. In practical scenarios, the Fermi velocity $v_F$, chemical potential $\mu$, and graphene's scattering rate $\hbar\Gamma$ are respectively set to be $10^6$ m/s, 400 meV, and 6.8 meV.[21]

In Eq. 2 and 3, the first term on the right-hand side corresponds to intraband transitions within graphene's conduction or valence band, and the second term corresponds to interband transitions between the valence and conduction bands. In the THz band where the photon energy $\hbar\omega$ is much less than $2\mu$ (= 800 meV), only the intraband transitions are allowed, and thus graphene behaves like metals exhibiting high conductivity of both longitudinal and Hall components. To design an integrated optical isolator for fiber-optic communication, we are especially interested in the wavelength near 1.55 μm, where graphene behaves like a semiconductor with a relatively small conductivity owning to the interband transitions for photon energy $\hbar\omega \geq 2\mu$ (= 800 meV). Under the condition of $T$ = 77 K and $B_0$ = 8.4 T, we evaluate graphene's longitudinal and Hall conductivities with the results shown in Fig. 2. One may notice that $\sigma_L$ exhibits a nontrivial value for photon energy higher than 0.8 eV, while $\sigma_H$ is nontrivial for photon energy near 0.8 eV. This stark contrast originates from the symmetry of the Landau energy levels and the band structure of graphene (see Part I of supplemental material[33]). It should also be noted that, graphene's Hall conductivity $\sigma_H$ depends strongly on temperature $T$ and magnetic field $B_0$. Either a high temperature or a weak magnetic field could lead to reduction of $\sigma_H$ (see Part II of supplemental material[33]).

To utilize the graphene's magneto-optical property, the photonic spin in integrated optical isolators has to be normal to the graphene sheet. Therefore, the spin of light has to be orthogonal to the propagation direction, which seemingly violates the common sense that light is a transverse wave. Actually, light propagating in free space and uniform media is indeed a transverse electromagnetic wave with the photonic spin in parallel with the light propagation direction. However, when light is confined to a subwavelength waveguide, it is no longer a pure transverse wave and its electric field carries a longitudinal component, which is caused by the total internal reflection at the boundaries of the waveguide. In this case, the optical spin is normal to the light propagation direction and changes its sign when the propagation direction is reversed.[28] Since the light propagation direction by definition determines the orbital angular momentum of the photons in the microring, now the light's spin and orbital parts of angular momentum are interrelated, which is referred to as "photonic spin–orbit coupling".[29]

By finite-element simulation, we conducted full-vector eigenmode analysis for a bent silicon waveguide as shown in Fig. 3(a), where the waveguide width and height are 0.75 μm and 0.25 μm respectively, and the bending radius is set to be 20 μm. A single TE-like fundamental mode is supported around the wavelength of 1.55 μm, with the spatial profile of its transverse component $E_\rho$ and longitudinal component $E_\varphi$ in Fig. 3(b). The commonly studied $E_\rho$ component is mainly confined to the waveguide with discontinuity at the silicon–air interfaces. The $E_\varphi$ component reaches maximum on the waveguide edges and zero near the center. It should be noted that, the $E_\rho$ and $E_\varphi$ components always have a constant phase difference of $\pm\pi/2$, in which the sign is determined by the direction of light propagation. As a result, the $E_\rho$ and $E_\varphi$ components form a right or left elliptically polarized field for counterpropagating light, which results in the photonic spin perpendicular to the light propagation direction.[30] This is fundamentally different from the case of light propagating in free space or uniform media, where the optical spin could only align to the propagation direction.

We further calculated the distribution of electromagnetic spin density $\mathbf{S} = \mathbf{S}_e + \mathbf{S}_m$, where $\mathbf{S}_e$ and $\mathbf{S}_m$ are respectively the electric and magnetic contribution of optical spin:[30]

$$\begin{cases} \mathbf{S}_e = \mathrm{Im}\left(\varepsilon \mathbf{E}^* \times \mathbf{E}\right)/4\omega \\ \mathbf{S}_m = \mathrm{Im}\left(\mu \mathbf{H}^* \times \mathbf{H}\right)/4\omega \end{cases} \quad (4)$$

In the above equations, $\mathbf{E}$ ($\mathbf{H}$) denotes the vectorial electric (magnetic) field of the optical mode. As electrons in the graphene interact mainly with the $\mathbf{E}$ component of light, we consider only the electric contribution $\mathbf{S}_e$ in this work. Figure 3(b) shows the calculated individual components of $\mathbf{E}$ ($E_\rho$ and $E_\varphi$) and the corresponding $\mathbf{S}_e$ for both forward and backward propagating light. It is obvious that $\mathbf{S}_e$ only contains the z component and changes its sign when the light propagation direction is reversed. This is caused by the opposite phase difference between the $E_\rho$ and $E_\varphi$ components as a natural consequence of time-reversal symmetry of light traveling in an optical waveguide. In our proposed device shown in Fig. 1, a 0.2-μm-wide graphene nanoribbon covers the inner top surface of the bent waveguide. The optical spin $\mathbf{S}_e$ is perpendicular to the graphene's carbon atom plane when light is propagating along the waveguide. Therefore, the graphene will experience opposite optical spin for counterpropagating light modes.

With graphene's conductivity tensor $\boldsymbol{\sigma}$, we calculated the effective refractive index $n_f$ ($n_b$) for the forward (backward) light mode by finite-element method in COMSOL.[34] By including the magnetically induced Hall conductivity as well as the photonic spin–orbit coupling effect, the graphene on top of the silicon waveguide causes a slight difference between $n_f$ and $n_b$, thus producing nonreciprocal transmission for light traveling in different directions. This nonreciprocal transmission effect can be enhanced further by resonance if an optical cavity is employed. Shown in Fig. 1(a), if we denote $r$ the coupling efficiency between the bus waveguide and the microring cavity, then the forward and backward light transmission spectrum can be expressed as:[35]

$$T_i = \frac{a_i^2 - 2ra_i\cos\theta_i + r^2}{1 - 2ra_i\cos\theta_i + r^2 a_i^2} \quad (i = f, b) \quad (5)$$

where $a_i = \exp[-2\pi R\, k_0\, \mathrm{Im}(n_i)]$ and $\theta_i = 2\pi R\, k_0\, \mathrm{Re}(n_i)$ are respectively the single-pass amplitude transmission and phase shift for the forward ($i = f$) and backward ($i = b$) propagating mode. The radius $R$ of the microring is 20 μm and $k_0$ denotes the light's wavevector in vacuum. As shown in Eq. 5, when $\cos\theta_i = 1$ is achieved, resonance occurs and the energy transmission $T_i$ is simplified as $(a_i - r)^2/(1 - ra_i)^2$. Now $T_i$ reaches zero under the so-called "critical coupling" condition when the coupling efficiency $r$ is equal to the single-pass amplitude transmission $a_i$.[35] As $a_f$ and $a_b$ can be obtained from the imaginary part of $n_f$ and $n_b$ with different values, large extinction ratio can be achieved near an optical resonance for counterpropagating light, as shown in the transmission spectra in Fig. 4. At temperature of 77 K and magnetic field of 8.4 T, a large extinction ratio of ~45.3 dB with an insertion loss of ~12.3 dB can be achieved from our proposed device at the wavelength of 1.552 μm. It should be noted that although graphene's magneto-optical effect can also lead to distinct real parts of $n_f$ and $n_b$, inducing a slight difference between $\theta_f$ and $\theta_b$ and causing the resonant wavelengths to shift, such a wavelength shift is too small when compared to the resonant linewidth of 0.12 nm and thus can be safely ignored.

In fact, the nonreciprocal transmission performance is dominated by graphene's Hall conductivity $\sigma_H$, which depends on the external magnetic field $B_0$ and temperature $T$ (see Part II of supplemental material[33]). We further calculated the refractive index difference ($\Delta n_{\text{eff}} = n_f - n_b$) of counterpropagating light at the wavelength of 1.55 μm with different values of $B_0$ and $T$. As shown in Fig. 5, at low temperature and strong magnetic field, graphene's Hall conductivity $\sigma_H$ is nontrivial, with both real and imaginary part of $\Delta n_{\text{eff}}$ easily observable. However, under weaker magnetic field $B_0$ or higher temperature $T$, graphene's Hall conductivity is reduced, causing $\Delta n_{\text{eff}}$ to approach zero. It should be noted that the oscillations of both Re($\Delta n_{\text{eff}}$) and Im($\Delta n_{\text{eff}}$) with varying $B_0$ inherit directly from that of $\sigma_H$ (see Part I of supplemental material[33]).

Compared with previous research on graphene-based optical isolators, our proposal exploits graphene's magneto-optical properties on an integrated platform. Moreover, most graphene-based isolators reported to date operate in the THz frequencies while our proposed device is designed to work in the conventional fiber-optic communication band near the wavelength of 1.55 μm. Actually, for those graphene-based nonreciprocal devices operating with an out-of-plane scheme, the light polarization rotation angle is proportional to the real part of the Hall conductivity $\sigma_H$.[20] Considering that the Faraday rotation is only a few degrees with a giant value of $\sigma_H$ in the THz regime, such an effect could hardly be detected for light of near-infrared wavelengths due to the considerably lower Hall conductivity. This difficulty has now been overcome with our design using an in-plane scheme where light propagates along (rather than perpendicularly to) the graphene sheet to significantly enhance the light–graphene interaction. As a result, we can obtain excellent nonreciprocal isolation performance despite graphene's much lower Hall conductivity in the communication band. Compared with conventional optical isolators utilizing magneto-optical garnets, our proposed device with significantly reduced footprint of $1.26 \times 10^3$ μm$^2$ can achieve excellent isolation performance (see Table I). Our devices are also easier to fabricate because of the natural compatibility of two-dimensional materials with the traditional semiconductor technology. Moreover, unlike the nonlinear optical isolators[36] which require high input power and a high-$Q$ cavity and suffer from a limitation due to dynamic reciprocity,[37] our design does not set any limitation to the power of the input light. On the other hand, the requirement of cryogenic temperatures and strong magnetic field remains challenging for practical applications in on-chip photonic systems presently. This requirement originates from the Landau energy of graphene's band structure. A possible way to overcome this difficulty would be strain engineering of graphene. There are experiments showing that a designed strain can induce strong gauge fields that effectively act as a uniform magnetic field,[38] which can also lead to giant Faraday rotation for realizing optical isolators.[39]

In conclusion, we have proposed a design of graphene/silicon integrated isolator by exploiting graphene's magneto-optical effect and photonic spin–orbit coupling in nanowaveguides. Such a device is optimized for operation in the fiber-optic communication band near the wavelength of 1.55 μm. With significantly reduced device footprint, good isolation performance, and graphene's CMOS compatibility, such devices based on two-dimensional materials have shown great promise in on-chip integration of photonic nonreciprocal devices.

This research was supported by project BME-p5-15 of the Shun Hing Institute of Advanced Engineering and Direct Grant for Research of the Faculty of Engineering, The Chinese University of Hong Kong. The authors acknowledge fruitful discussion with Prof. Jian-Bin Xu, Department of Electronic Engineering, The Chinese University of Hong Kong.

**Figures**

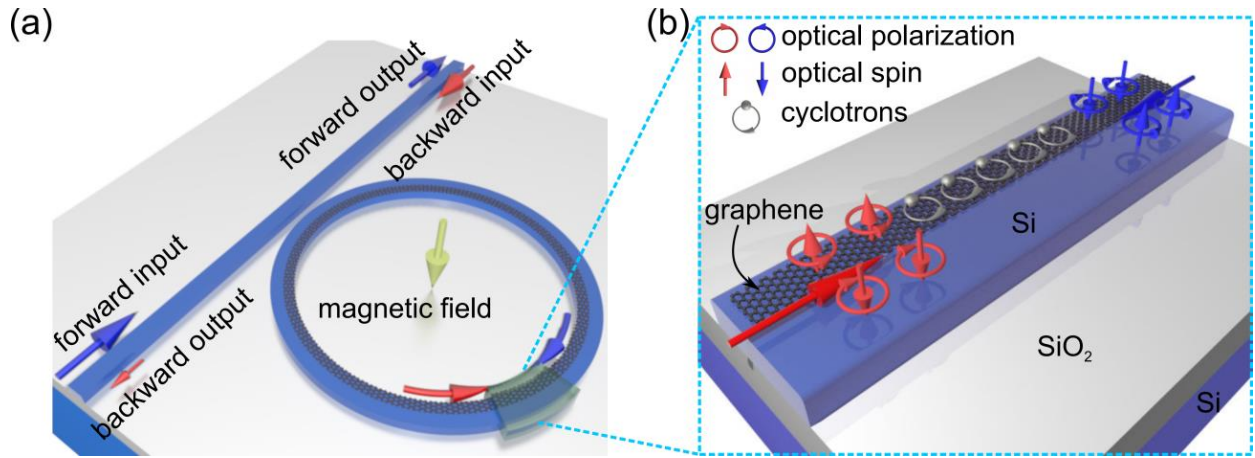

FIG. 1. (a) Schematic of the proposed hybrid graphene/silicon integrated optical isolator, which consists of a bus waveguide in close proximity of a microring cavity. The inner top surface of the microring is covered by a graphene nanoribbon, which under an external magnetic field produces nonreciprocal transmission for light propagating in different directions. (b) A zoomed section of the microring showing mechanism of the nonreciprocal optical transmission. The magnetically induced cyclotrons in the graphene sheet experiences opposite photonic spins for counterpropagating light, thus producing different optical response to forward and backward light modes.

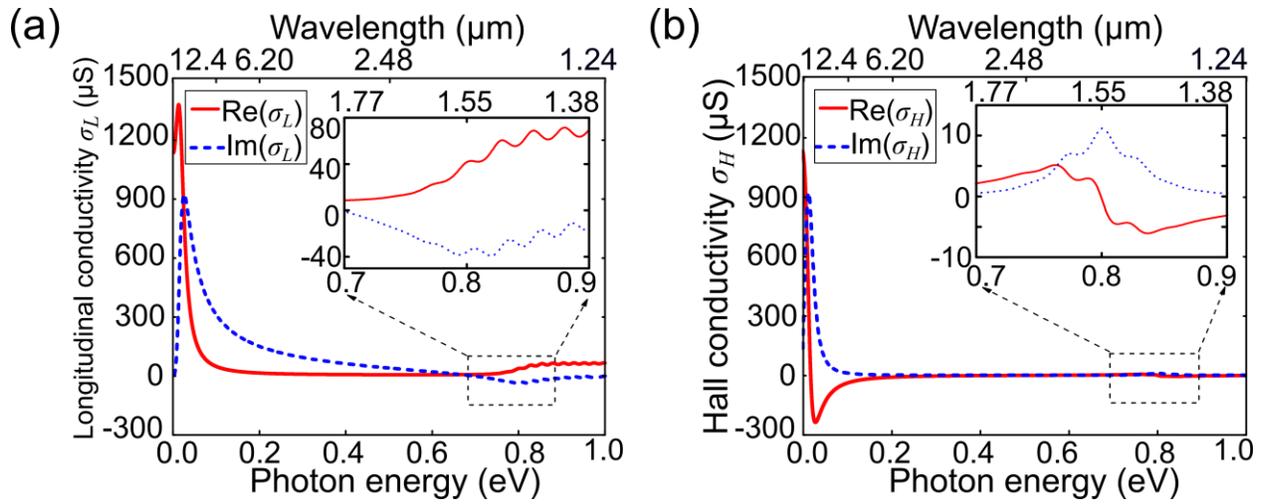

FIG. 2. (a, b) Optical spectra of longitudinal ($\sigma_L$) and Hall ($\sigma_H$) conductivities of graphene under an external magnetic field of 8.4 T and temperature of 77 K. The Hall conductivity shown in (b) takes nonzero values in two predominant spectral regimes: one in the THz band owing to electron's intraband transitions and the one in the near-infrared band owing to electron's interband transitions.

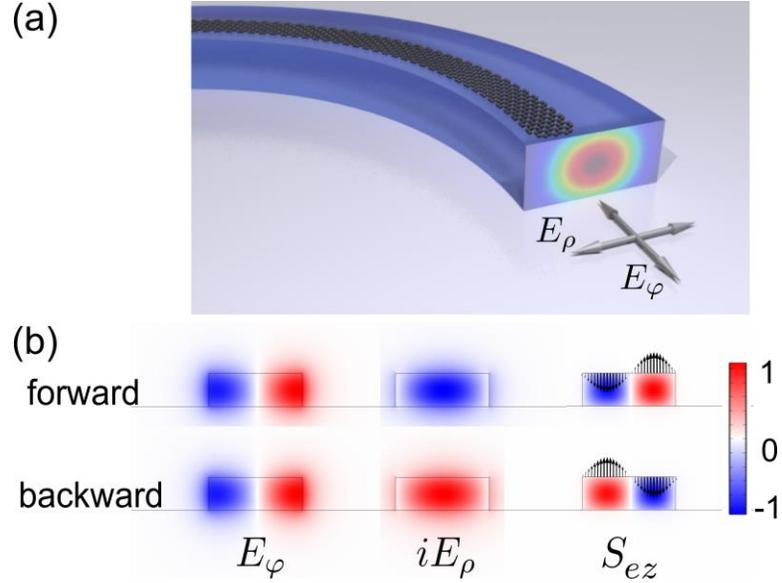

FIG. 3. (a) Sketch showing the fundamental TE-like mode of a bent silicon waveguide with graphene atop. The two major electric field components $E_\rho$ and $E_\varphi$ are indicated by the arrows. (b) Spatial distribution of $E_\rho$, $E_\varphi$, and the electric spin $S_{ez}$ for both forward and backward propagating light modes. The arrows superimposed onto the profile of $S_{ez}$ indicate the direction and strength of $\mathbf{S}_e$ at the interface between the graphene and the silicon waveguide. Note that the phase difference of $\pm\pi/2$ between the major $E_\rho$ and $E_\varphi$ components respectively for the forward and backward light produces opposite spatial profile of optical spin for interaction with the graphene.

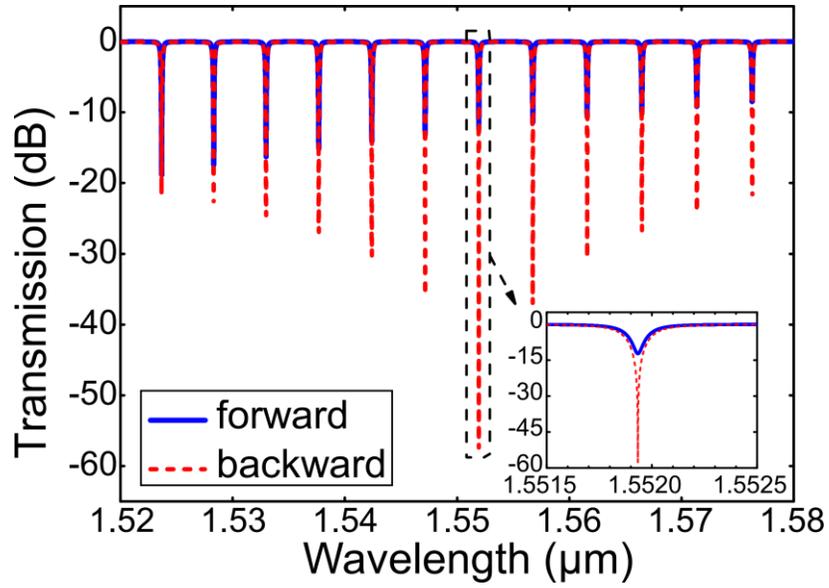

FIG. 4. Calculated forward and backward light transmission spectra of the proposed device under the condition of temperature 77 K and magnetic field 8.4 T. The resonance at 1.552 μm exhibits a high extinction ratio of ~45.3 dB with a low insertion loss of ~12.3 dB.

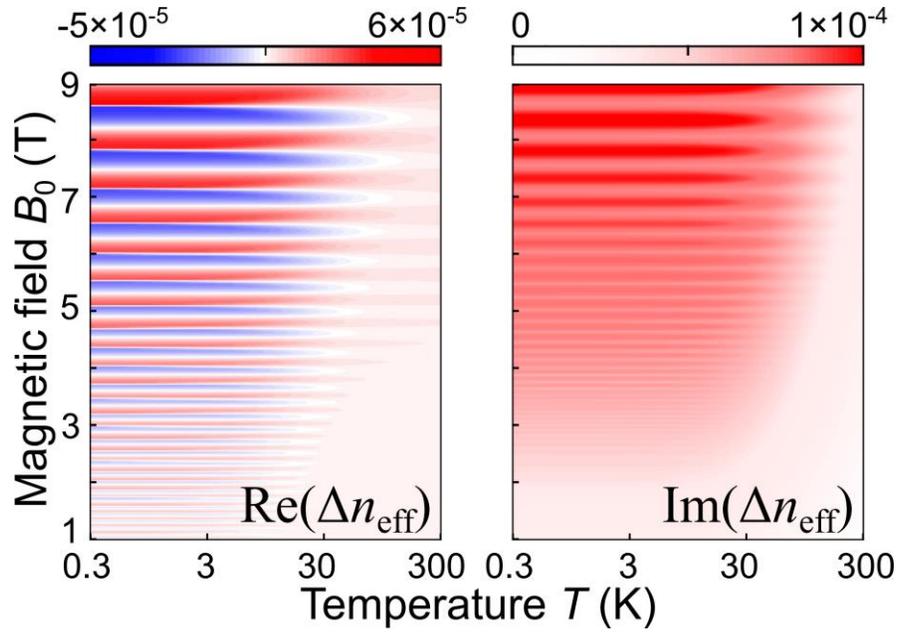

FIG. 5. Calculated real and imaginary part of the effective refractive index difference ($\Delta n_{\text{eff}}$) of the counterpropagating light at the wavelength of 1.55 μm under different temperature and magnetic field. Either higher temperature $T$ or weaker magnetic field $B_0$ can reduce $\Delta n_{\text{eff}}$, causing weaker nonreciprocal effects.

**Table**

TABLE I. Comparison of isolation performance and device footprint with other integrated optical isolators.

| Reference | Isolation performance | | Device size ($\mu m^2$) |
|---|---|---|---|
| | Extinction ratio (dB) | Insertion loss (dB) | |
| Ref. 2 | 19.5 | 18 | ~ $2.40 \times 10^5$ |
| Ref. 5 | 9 | 48 | ~ $2.54 \times 10^6$ |
| Ref. 7 | 25 | 9 | > $1.84 \times 10^6$ |
| Ref. 8 | 21 | 8 | ~ $1.20 \times 10^6$ |
| This proposal | 45.3 | 12.3 | ~ $1.26 \times 10^3$ |

**References**

1. D. Jalas, A. Petrov, M. Eich, W. Freude, S. Fan, Z. Yu, R. Baets, M. Popovic, A. Melloni, J. D. Joannopoulos, M. Vanwolleghem, C. R. Doerr, and H. Renner, Nat. Photonics **7**(8), 579–582 (2013).
2. L. Bi, J. Hu, P. Jiang, D. H. Kim, G. F. Dionne, L. C. Kimerling, and C. A. Ross, Nat. Photonics **5**(12), 758–762 (2011).
3. K. Petermann, IEEE J. Sel. Top. Quantum Electron. **1**(2), 480–489 (1995).
4. B. J. H. Stadler and T. Mizumoto, IEEE Photonics J. **6**(1), 1–15 (2014).
5. M.-C. Tien, T. Mizumoto, P. Pintus, H. Kromer, and J. E. Bowers, Opt. Express **19**(12), 11740–11745 (2011).
6. S. Yuya and M. Tetsuya, Sci. Technol. Adv. Mater. **15**(1), 014602 (2014).
7. S. Ghosh, S. Keyvavinia, W. Van Roy, T. Mizumoto, G. Roelkens, and R. Baets, Opt. Express **20**(2), 1839–1848 (2012).
8. T. Mizumoto, R. Takei, and Y. Shoji, IEEE J. Quantum Electron. **48**(2), 252–260 (2012).
9. T. Boudiar, B. Payet-Gervy, M. F. Blanc-Mignon, J. J. Rousseau, M. Le Berre, and H. Joisten, J. Magn. Magn. Mater. **284**, 77–85 (2004).
10. H.-S. Kim, L. Bi, G. F. Dionne, and C. A. Ross, Appl. Phys. Lett. **93**(9), 092506 (2008).
11. L. Bi, J. Hu, P. Jiang, H. Kim, D. Kim, M. Onbasli, G. Dionne, and C. Ross, Materials **6**(11), 5094–5117 (2013).
12. X. Gan, R.-J. Shiue, Y. Gao, I. Meric, T. F. Heinz, K. Shepard, J. Hone, S. Assefa, and D. Englund, Nat. Photonics **7**(11), 883–887 (2013).
13. X. Wang, Z. Cheng, K. Xu, H. K. Tsang, and J.-B. Xu, Nat. Photonics **7**(11), 888–891 (2013).
14. M. Liu, X. Yin, E. Ulin-Avila, B. Geng, T. Zentgraf, L. Ju, F. Wang, and X. Zhang, Nature **474**(7349), 64–67 (2011).
15. C. T. Phare, Y.-H. Daniel Lee, J. Cardenas, and M. Lipson, Nat. Photonics **9**(8), 511–514 (2015).
16. Y. D. Kim, H. Kim, Y. Cho, J. H. Ryoo, C.-H. Park, P. Kim, Y. S. Kim, S. Lee, Y. Li, S.-N. Park, Y. Shim Yoo, D. Yoon, V. E. Dorgan, E. Pop, T. F. Heinz, J. Hone, S.-H. Chun, H. Cheong, S. W. Lee, M.-H. Bae, and Y. D. Park, Nat. Nanotechnol. **10**(8), 676–681 (2015).
17. K. S. Novoselov, A. K. Geim, S. V. Morozov, D. Jiang, M. I. Katsnelson, I. V. Grigorieva, S. V. Dubonos, and A. A. Firsov, Nature **438**(7065), 197–200 (2005).
18. A. H. Castro Neto, F. Guinea, N. M. R. Peres, K. S. Novoselov, and A. K. Geim, Rev. Mod. Phys. **81**(1), 109–162 (2009).
19. K. S. Novoselov, V. I. Falko, L. Colombo, P. R. Gellert, M. G. Schwab, and K. Kim, Nature **490**(7419), 192–200 (2012).
20. I. Crassee, J. Levallois, A. L. Walter, M. Ostler, A. Bostwick, E. Rotenberg, T. Seyller, D. van der Marel, and A. B. Kuzmenko, Nat. Phys. **7**(1), 48–51 (2011).
21. R. Shimano, G. Yumoto, J. Y. Yoo, R. Matsunaga, S. Tanabe, H. Hibino, T. Morimoto, and H. Aoki, Nat. Commun. **4**, 1841 (2013).
22. I. Crassee, M. Orlita, M. Potemski, A. L. Walter, M. Ostler, T. Seyller, I. Gaponenko, J. Chen, and A. B. Kuzmenko, Nano Lett. **12**(5), 2470–2474 (2012).
23. N. Ubrig, I. Crassee, J. Levallois, I. O. Nedoliuk, F. Fromm, M. Kaiser, T. Seyller, and A. B. Kuzmenko, Opt. Express **21**(21), 24736–24741 (2013).
24. A. Ferreira, J. Viana-Gomes, Y. V. Bludov, V. Pereira, N. M. R. Peres, and A. H. Castro Neto, Phys. Rev. B **84**(23), 235410 (2011).
25. H. Da and C.-W. Qiu, Appl. Phys. Lett. **100**(24), 241106 (2012).
26. V. S. Liberman and B. Y. Zel'dovich, Phys. Rev. A **46**(8), 5199–5207 (1992).
27. K. Y. Bliokh, F. J. Rodriguez-Fortuno, F. Nori, and A. V. Zayats, Nat. Photonics **9**(12), 796–808 (2015).
28. R. Mitsch, C. Sayrin, B. Albrecht, P. Schneeweiss, and A. Rauschenbeutel, Nat. Commun. **5**, 5713 (2014).
29. J. Petersen, J. Volz, and A. Rauschenbeutel, Science **346**(6205), 67–71 (2014).
30. K. Y. Bliokh, A. Y. Bekshaev, and F. Nori, Nat. Commun. **5**, 3300 (2014).
31. V. P. Gusynin and S. G. Sharapov, Phys. Rev. B **73**(24), 245411 (2006).

32 K. S. Novoselov, Z. Jiang, Y. Zhang, S. V. Morozov, H. L. Stormer, U. Zeitler, J. C. Maan, G. S. Boebinger, P. Kim, and A. K. Geim, Science **315**(5817), 1379–1379 (2007).
33 See supplemental material at [URL will be inserted by AIP] for discussion of graphene's Hall conductivity
34 https://www.comsol.com/
35 A. Yariv, IEEE Photonics Technol. Lett. **14**(4), 483–485 (2002).
36 L. Fan, J. Wang, L. T. Varghese, H. Shen, B. Niu, Y. Xuan, A. M. Weiner, and M. Qi, Science **335**(6067), 447–450 (2012).
37 Y. Shi, Z. Yu, and S. Fan, Nat. Photonics **9**(6), 388–392 (2015).
38 F. Guinea, M. I. Katsnelson, and A. K. Geim, Nat. Phys. **6**(1), 30–33 (2010).
39 J. C. Martinez, M. B. A. Jalil, and S. G. Tan, Opt. Lett. **37**(15), 3237–3239 (2012).

32 K. S. Novoselov, Z. Jiang, Y. Zhang, S. V. Morozov, H. L. Stormer, U. Zeitler, J. C. Maan, G. S. Boebinger, P. Kim, and A. K. Geim, Science **315**(5817), 1379–1379 (2007).
33 See supplemental material at [URL will be inserted by AIP] for discussion of graphene's Hall conductivity
34 https://www.comsol.com/
35 A. Yariv, IEEE Photonics Technol. Lett. **14**(4), 483–485 (2002).
36 L. Fan, J. Wang, L. T. Varghese, H. Shen, B. Niu, Y. Xuan, A. M. Weiner, and M. Qi, Science **335**(6067), 447–450 (2012).
37 Y. Shi, Z. Yu, and S. Fan, Nat. Photonics **9**(6), 388–392 (2015).
38 F. Guinea, M. I. Katsnelson, and A. K. Geim, Nat. Phys. **6**(1), 30–33 (2010).
39 J. C. Martinez, M. B. A. Jalil, and S. G. Tan, Opt. Lett. **37**(15), 3237–3239 (2012).

Supplemental material for "Hybrid graphene/silicon integrated optical isolators with photonic spin–orbit interaction"

Jingwen Ma,[1,2] Xiang Xi,[1] Zejie Yu,[1] and Xiankai Sun[1,2,a]

[1]Department of Electronic Engineering, The Chinese University of Hong Kong, Shatin, New Territories, Hong Kong
[2]Shun Hing Institute of Advanced Engineering, The Chinese University of Hong Kong, Shatin, New Territories, Hong Kong

In this supplemental material we analyze graphene's electronic band structure and Landau energy levels to explain the different features of the longitudinal ($\sigma_L$) and Hall ($\sigma_H$) conductivities in the near-infrared regime. The effects of varying external magnetic field $B_0$ and temperature $T$ on graphene's Hall conductivity $\sigma_H$ are also discussed.

**Part I: Explanation of graphene's longitudinal and Hall conductivities in the near-infrared regime**

Graphene's conductivity can be expressed using the Kubo method as follows:

$$\sigma_L = \frac{e^2 v_F^2 |eB_0|(\omega - j2\Gamma)\hbar}{j\pi}$$
$$\times \sum_{n=0}^{\infty} \left[ \frac{(f(E_{i+1}) - f(E_i)) + (f(-E_i) - f(-E_{i+1}))}{(E_{i+1} - E_i)^2 - (\omega - j2\Gamma)^2 \hbar^2} \times \frac{1 - \Delta^2/E_i E_{i+1}}{E_{i+1} - E_i} \right.$$
$$\left. + \frac{(f(E_{i+1}) - f(-E_i)) + (f(E_i) - f(-E_{i+1}))}{(E_{i+1} + E_i)^2 - (\omega - j2\Gamma)^2 \hbar^2} \times \frac{1 + \Delta^2/E_i E_{i+1}}{E_{i+1} + E_i} \right] \quad (S1)$$

$$\sigma_H = \frac{e^2 v_F^2 eB_0}{\pi} \times \sum_{n=0}^{\infty} \left[ (f(E_{i+1}) - f(-E_i)) - (f(E_i) - f(-E_{i+1})) \right]$$
$$\times \left[ \frac{1 - \Delta^2/E_i E_{i+1}}{(E_{i+1} - E_i)^2 - (\omega - j2\Gamma)^2 \hbar^2} + \frac{1 + \Delta^2/E_i E_{i+1}}{(E_{i+1} + E_i)^2 - (\omega - j2\Gamma)^2 \hbar^2} \right] \quad (S2)$$

In the above equations, the first term on the right-hand side corresponds to intraband transitions within graphene's conduction ($E_i \to E_{i+1}$) or valence band ($-E_{i+1} \to -E_i$), and the second term corresponds to interband transitions between the valence and conduction bands ($-E_i \to E_{i+1}$ and $-E_{i+1} \to E_i$). Owing to the symmetry of the Landau energy levels and the band structure of graphene, most interband transitions occur in pairs. For example, both transitions ($-E_i \to E_{i+1}$) and ($-E_{i+1} \to E_i$) can be excited by photons of energy $\hbar\omega = E_i + E_{i+1}$. The only exception is the process ($-E_N \to E_{N+1}$), where $N$ is the integer satisfying $E_{N+1} > \mu > E_N$. As shown in Fig. S1(c) and S1(d), ($-E_N \to E_{N+1}$) occurs when $\hbar\omega$ equals $E_N + E_{N+1}$, but its counterpart process ($-E_{N+1} \to$

---

[a]Author to whom correspondence should be addressed. Electronic mail: xksun@cuhk.edu.hk

$E_N$) is forbidden because the energy level $E_N$ is occupied. With careful inspection, one can find out the contribution of those interband transitions to the optical conductivities. More specifically, the term $(f(E_{i+1}) - f(-E_i)) + (f(E_i) - f(-E_{i+1}))$ in Eq. S1 and the term $(f(E_{i+1}) - f(-E_i)) - (f(E_i) - f(-E_{i+1}))$ in Eq. S2 have dictated that the transition pairs $(-E_i \to E_{i+1}$ and $-E_{i+1} \to E_i, i > N)$ contribute constructively to $\sigma_L$ and destructively to $\sigma_H$. This explains the stark contrast between $\sigma_L$ and $\sigma_H$ shown in insets of Fig. S1(a) and S1(b): The oscillation of $\sigma_L$ rising near 0.8 eV results from many nonequivalent interband transitions that contribute to the optical spectral weight. For $\sigma_H$, the contribution from transition pairs $(-E_i \to E_{i+1}$ and $-E_{i+1} \to E_i, i > N)$ all cancels to zero, leaving only the lowest interband transition $(-E_N \to E_{N+1})$ to play a role.

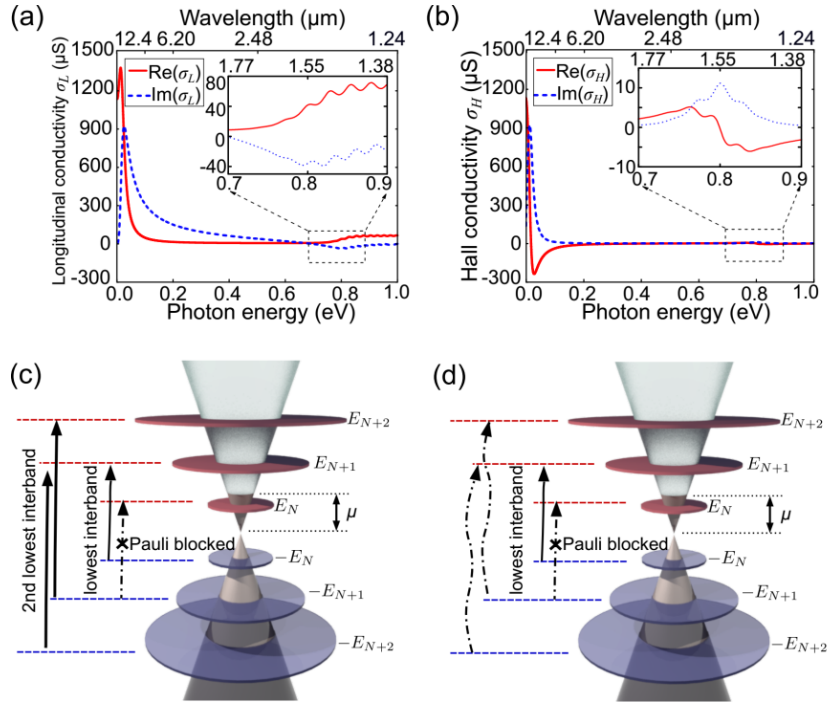

FIG. S1. (a, b) Optical spectra of longitudinal ($\sigma_L$) and Hall ($\sigma_H$) conductivities of graphene under an external magnetic field of 8.4 T and temperature of 77 K. The Hall conductivity shown in (b) takes nonzero values in two predominant spectral regimes: one in the THz band owing to electron's intraband transitions and the one in the near-infrared band owing to electron's interband transitions. (c, d) Graphene's electronic band structure and Landau energy levels in an external magnetic field for explaining the features of $\sigma_L$ and $\sigma_H$ in the near-infrared regime. Most of electron's interband transitions like $(-E_i \to E_{i+1}$ and $-E_{i+1} \to E_i, i > N)$ occur in pairs, contributing constructively to $\sigma_L$ (c) and destructively to $\sigma_H$ (d). Here $E_N$ is the $N$-th Landau level satisfying $E_{N+1} > \mu > E_N$. The only single transition where $(-E_N \to E_{N+1})$ is allowed but $(-E_{N+1} \to E_N)$ is Pauli blocked gives rise to the nonzero $\sigma_H$.

## Part II: Effects of varying external magnetic field and temperature on graphene's Hall conductivity

It is also interesting to investigate the effects of weaker magnetic field $B_0$ and higher temperature $T$ on the Hall conductivity $\sigma_H$ of graphene. The magnetic field $B_0$ affects $\sigma_H$ in two aspects. Firstly, the peak optical frequency is related to $B_0$. As shown in Fig. S2(a), the light frequency satisfying $\hbar\omega = E_N + E_{N+1}$ shifts to the blue side as $B_0$ increases. This is mostly because the Landau energy levels $E_i = \left(\Delta^2 + 2i|eB_0|\hbar v_F^2\right)^{1/2}$ are determined by $|B_0|$. The oscillating behavior distinct at low temperatures (e.g., $T = 77$ K) is a direct result of the quantized Landau energy levels, a phenomenon with the same origin as the famous De Haas–van Alphen effect.[1] Secondly, stronger magnetic field $B_0$ induces larger peak value of $|\sigma_H|$. As shown in Fig. S2(b), $B_0$ of 1.3 T, 4.1 T, and 8.4 T are selected to ensure identical peak optical wavelength around 1.55 μm. It is obvious that both the real and imaginary parts of $\sigma_H$ pick up higher values with increasing $B_0$. Rise of temperature weakens graphene's magneto-optical effect due to the broadening of the Fermi–Dirac distribution at the Fermi surface. It can be inferred mathematically that the term $(f(E_{i+1}) - f(-E_i)) - (f(E_i) - f(-E_{i+1}))$ is no longer zero for $i > N$ at higher temperatures, indicating that the contribution of transition pairs ($-E_i \rightarrow E_{i+1}$ and $-E_{i+1} \rightarrow E_i$, $i > N$) does not cancel to zero. This is further supported by the spectrum of $|\sigma_H|$ at $T = 300$ K in Fig. S2(a), which shows multiple peaks in the wavelength range of 1.2 to 1.9 μm. On the other hand, increase of temperature leads to an overall reduction of the conductivity, due to the smaller value of the term $(f(E_{N+1}) - f(-E_N)) - (f(E_N) - f(-E_{N+1}))$.

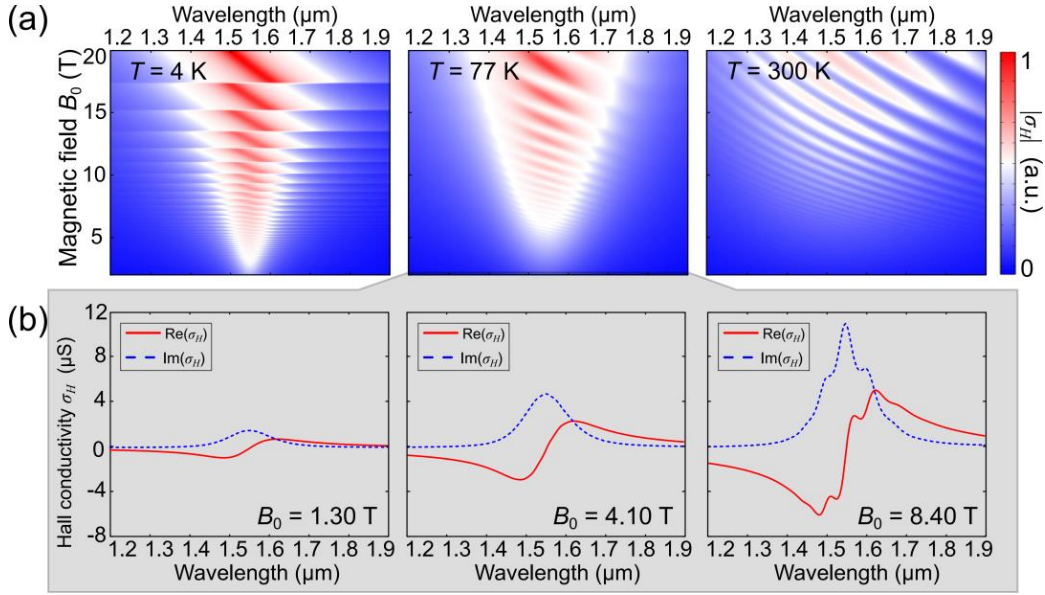

FIG. S2. (a) Computed optical spectra of graphene's Hall conductivity $|\sigma_H|$ as a function of the external magnetic field $B_0$. The cases under three representative temperatures $T = 4$ K, 77 K, and 300 K are plotted. (b) Evolution of graphene's Hall conductivity [both Re($\sigma_H$) and Im($\sigma_H$)] with a varying external magnetic field $B_0$ and a fixed temperature of 77 K. The three $B_0$ values 1.3 T, 4.1 T, and 8.4 T are chosen such that the Hall conductivity peaks at around 1.55 μm.

# References

[1] J. P. Eisenstein, H. L. Stormer, V. Narayanamurti, A. Y. Cho, A. C. Gossard, and C. W. Tu, Phys. Rev. Lett. **55**(8), 875–878 (1985).